# FPGA-Optimized Hardware Accelerator for Fast Fourier Transform and Singular Value Decomposition in AI

Hong Ding
*School of Electrical Engineering and Artificial Intelligence*
*Xiamen University Malaysia*
Sepang, Malaysia
EEE2109269@xmu.edu.my

Chia Chao Kang
*School of Electrical Engineering and Artificial Intelligence*
*Xiamen University Malaysia*
Sepang, Malaysia
chiachao.kang@xmu.edu.my

SuYang Xi
*School of Electrical Engineering and Artificial Intelligence*
*Xiamen University Malaysia*
Sepang, Malaysia
EEE2109299@xmu.edu.my

Zehang Liu
*School of Computer and Data Science*
*Xiamen University Malaysia*
Sepang, Malaysia
CST2309159@xmu.edu.my

Xuan Zhang
*School of Computer and Data Science*
*Xiamen University Malaysia*
Sepang, Malaysia
CST2309183@xmu.edu.my

Yi Ding
*School of Economics and Management*
*Xiamen University Malaysia*
Sepang, Malaysia
JRN2109439@xmu.edu.my

*Abstract*—This research introduces an FPGA-based hardware accelerator to optimize the Singular Value Decomposition (SVD) and Fast Fourier transform (FFT) operations in AI models. The proposed design aims to improve processing speed and reduce computational latency. Through experiments, we validate the performance benefits of the hardware accelerator and show how well it handles FFT and SVD operations. With its strong security and durability, the accelerator design achieves significant speedups over software implementations, thanks to its modules for data flow control, watermark embedding, FFT, and SVD.

## I. Introduction

Recently, artificial intelligence has made tremendous progress in image processing and artistic creation thanks to the quick development of generative AI and diffusion models [1]. Specifically, throughout different artist communities, text-to-image diffusion models have demonstrated remarkable capacity to mimic artist styles. This incredibly accurate and efficient generating power has changed the art market and sparked worries about safeguarding artists' uniqueness. Our earlier study suggested an AI watermarking approach based on the Fast Fourier transform (FFT) and Singular Value Decomposition (SVD) to address this. However, the computing speed and efficiency of existing software implementation approaches are becoming increasingly inadequate as data volumes continue to rise.

To address this obstacle, the current work introduces a hardware accelerator using FPGA technology, specifically intended to enhance the efficiency of Fast Fourier Transform (FFT) and Singular Value Decomposition (SVD) computations in artificial intelligence models. An FPGA, also known as a Field-Programmable Gate Array, is a very suitable option for enhancing computing performance. This is mostly due to its remarkable multitasking capabilities and its favorable low latency properties. Applying the Fast Fourier Transform (FFT) and Singular Value Decomposition (SVD) algorithms on a Field-Programmable Gate Array (FPGA) may greatly improve the speed and efficiency of image processing, hence facilitating the wider adoption of artificial intelligence (AI) applications.

Singular Value Decomposition (SVD) and the Fast Fourier Transform (FFT) are crucial in the processing and interpretation of image data [2][3]. The Fast Fourier Transform (FFT) is a mathematical procedure that converts pictures from the time domain to the frequency domain, therefore revealing hidden information. On the other hand, Singular Value Decomposition (SVD) is a method used to break down matrices in order to get the most significant eigenvectors of images. These techniques are often used for extracting features, compressing the dimensions of photographs, and applying digital watermarks. Previously, the implementation of this model in software and encountered issues with long execution times and high-performance demands. The current hardware accelerator aims to address these problems.

In this paper, it presented the design and implement an FPGA-based hardware accelerator to optimize and accelerate FFT and SVD operations. Experimental results demonstrate the significant improvements in processing speed and efficiency provided by this hardware accelerator, offering robust technical support for protecting artists' originality and enabling large-scale AI applications.

## II. Literature Review

The use of Fast Fourier transform (FFT) and Singular Value Decomposition (SVD) in digital watermarking technology has advanced significantly with the quick growth of AI, especially in the area of picture synthesis and processing. The use of FFT in CNNs and DNNs was examined in Reference [4][5], which also included significant advancements for autonomous optimization during training. Reference [6] proposed integrating masks, FFT, and Lorenz attractors to improve digital security. As reference [7] discussed digital watermarking techniques and their applications, FFT can be used in this context. Reference [8] emphasized the threat that diffusion models bring to the art industry and discussed how AI-based solutions like Glaze are essential to safeguarding artists. Strong watermark embedding and extraction capabilities are provided by SVD, which has been extensively used in digital watermarking because to its capacity to break down images into unique vectors and values. Mohanarathinam et al. (2020) classified digital watermarking approaches in Reference [5], emphasizing the usefulness of SVD-based techniques in guaranteeing data security and integrity. Their review suggested that combining SVD with FFT could significantly enhance the robustness of watermarking solutions.

In our earlier work, we found that when executing this watermarking model, long execution durations and frequent computer crashes were caused by the enormous image size, inefficient software implementation, and high computational complexity of the model. This problem made clear how urgently high-speed and efficient processing of models with large computational complexity is needed. FPGA (Field-Programmable Gate Array) technology offers significant advantages in parallel processing and low latency, making it an ideal choice for accelerating computational tasks. Through data transfer optimization and parallel computing unit execution, Miaoxin Wang et al. demonstrated considerable gains in computational efficiency in Reference [9], where they suggested an FPGA-based accelerator enabling CNNs with configurable kernel sizes. The accelerator achieved a reduction in overall latency by up to 65%. In Reference [10], Ali Arshad Nasir et al. demonstrated a flexible and efficient FPGA accelerator that can handle a range of lightweight and large-scale CNNs. On the Intel Arria 10 FPGA, the study showed excellent resource use and notable performance benefits. Certain performance improvements of several times to tens of times can be achieved with these FPGA-based hardware accelerators, depending on the application and optimization strategies employed. In conclusion, SVD and FFT integration in digital watermarking, along with FPGA-based hardware acceleration, offer a practical approach to safeguarding digital art in the AI age. The examined literature demonstrates how these developments can enhance the efficacy, security, and resilience of watermarking systems, safeguarding creative integrity in the digital era.

## III. METHODOLOGY

### 3.1 Fast Fourier Transform

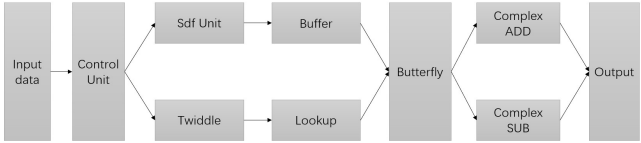

Fig. 1. FFT implementation logic diagram on FPGA

### 3.1.1 Definition of Fast Fourier Transform

The Fast Fourier Transform (FFT) is a very fast technique used for calculating the Discrete Fourier Transform (DFT) and its inverse. The Fast Fourier Transform (FFT) algorithm decreases the computational cost of calculating the Discrete Fourier Transform (DFT) from O($N^2$) to O(NlogN), where N represents the number of data points in the collection. FFT is very advantageous for handling extensive datasets in several fields, including signal processing, picture analysis, and audio processing, because to its substantial decrease in computing complexity. The Fast Fourier Transform (FFT) algorithm breaks down a signal into its individual sinusoidal components, enabling the examination of the signal's frequency spectrum.

A Fourier series is a continuous, periodic function created by a summation of harmonically related sinusoidal functions. It has several different, but equivalent, forms, shown here as partial sums. Determine the function $s_N(x)$ as follow: Fourier series, amplitude-phase form

$$s_N(x) = D_0 + \sum_{n=1}^{N} D_0 \cos\left(2\pi \frac{n}{p} x - \varphi_n\right) \quad (1)$$

Fourier series, sine-cosine form

$$s_N(x) = A_0 + \sum_{n=1}^{N} \left(A_n \cos\left(2\pi \frac{n}{p} x\right) + B_n \sin\left(2\pi \frac{n}{p} x\right)\right) \quad (2)$$

Fourier series, exponential form

$$s_N(x) = A_0 + \sum_{n=-N}^{N} C_n e^{i2\pi \frac{n}{p} x} \quad (3)$$

For the sine-cosine form, typically the coefficients are determined by frequency/harmonic analysis of a given real-valued function $s(x)$, and x represents time:

$$A_0 = \frac{1}{P} \int_P s(x)\, dx \quad (4)$$

$$A_n = \frac{2}{P} \int_P s(x) \cos\left(2\pi \frac{n}{p} x\right) dx \quad \text{for n} \geq 1 \quad (5)$$

$$B_n = \frac{2}{P} \int_P s(x) \sin\left(2\pi \frac{n}{p} x\right) dx \quad \text{for n} \geq 1 \quad (6)$$

### 3.1.3 Fourier Transform

The Fourier Transform is a mathematical method used to transform a signal from the time domain (or spatial domain) to the frequency domain. It enables us to express a function as a composition of sinusoidal functions, with each one corresponding to a distinct frequency. The Fourier Transform is very advantageous in the fields of signal processing, image analysis, and physics. The Fourier Transform is obtained by integrating complex numbers into the Fourier Series and extending it to the limit when the signal time approaches infinity. This procedure entails the conversion of the discrete coefficients $A_n$ into a continuous function $F(k)$ and the transformation of the sum into an integral. Therefore, The Fourier Transform of a function $f(x)$ is defined as:

$$f(x) = \int_{-\infty}^{\infty} F(x)\, e^{2\pi i k x} dk \quad (7)$$

$$F(x) = \int_{-\infty}^{\infty} f(x)\, e^{-2\pi i k x} dx \quad (8)$$

### 3.1.4 Fast Fractional Fourier Transform

For any integer N, the FFT is an efficient algorithm to compute the Discrete Fourier Transform (DFT) of a sequence of N complex numbers. The DFT of a sequence x[n] is denoted by X[k] and defined by

$$X[k] = \sum_{n=0}^{N-1} x[n] e^{-i\frac{2\pi}{N} kn} \quad for\ k = 0, 1 \ldots\ldots \quad (9)$$

A fundamental component of the FFT is the butterfly operation, which combines pairs of data points using twiddle factors (complex exponentials). The butterfly operation can be represented as:

$$X[k] = x[k] + W_N^m \mathrm{x}\left[k + \frac{N}{2}\right] \quad (10)$$

$$X\left[k + \frac{N}{2}\right] = x[k] - W_N^m \mathrm{x}\left[k + \frac{N}{2}\right] \quad (11)$$

Where $W_N^m = e^{-i\frac{2\pi}{N}m}$ is the twiddle factor, and N is the size of the FFT.

The Radix-2 FFT algorithm, one of the most common implementations, processes the input sequence in stages, splitting it into even-indexed and odd-indexed elements. Each stage performs butterfly operations, combining results to produce the final DFT.

### 3.1.5 Hardware Description:

This FPGA accelerator design utilizes numerous Single-path Delay Feedback (SDF) units to analyze input data, enabling the Fast Fourier Transform (FFT) and producing frequency domain output. The units consist of many SdfUnit modules and one SdfUnit2 module, which are coupled in a series to do hierarchical FFT processing. The research employs a cascade methodology to achieve Fast Fourier Transform (FFT) computation. Multiple Synchronous Data Flow (SDF) units are used, where each unit is allocated the responsibility of processing a distinct segment of the FFT computation and communicating the outcomes to the succeeding unit. The cascade structure effectively decomposes the complex FFT computation, simplifying the implementation of each component and improving the overall processing efficiency and speed.

Every SdfUnit carries out a certain degree of butterfly computing, which is the fundamental operation of the FFT algorithm. This calculation involves combining and recombining data points to produce the representation of the frequency domain. Each SdfUnit is equipped with a Delay Buffer to handle the data dependencies and time constraints of the butterfly calculation, guaranteeing accurate synchronization of data across multiple phases. Delay buffer units facilitate the synchronization of data flow between stages, thereby eliminating any instances of data collisions or losses, hence ensuring uninterrupted and seamless processing.

The SdfUnit2 module is a variant of the SDF unit, specifically designed to handle particular twiddle factor resolutions. Its parameters and ports are similar to those of the SdfUnit but differ in implementation details to meet the specific needs of the final stage. Twiddle factors are an essential part of FFT computation, corresponding to the exponential functions of complex numbers, used to convert time-domain data to the frequency domain. Each twiddle factor calculation requires high-precision multiplication to ensure accurate frequency domain conversion.

The data first enters the SdfUnit for the purpose of doing preliminary butterfly calculation. Every SdfUnit divides its incoming data into two segments, implements the FFT butterfly algorithm, calculates the output, and then transfers it to the next SdfUnit for further processing. To achieve correct frequency domain conversion, it is necessary to multiply the output of each step by specified twiddle factors. For efficient management of these intricate calculations, each SdfUnit is equipped with numerous arithmetic units and control logic to synchronize the execution of different operations.

Delay Buffer units play a crucial role in the FFT computing process by ensuring smooth data flow between processing units and ensuring precise timing and synchronization. Ultimately, after all rounds of processing have been finished, the resultant frequency domain data may be used for further signal analysis and processing. The structured and modular design improves computational efficiency and provides substantial adaptability and expandability in practical scenarios.

### 3.2 Singular Value Decomposition

### 3.2.1 Definition of Singular Value Decomposition

The use of Singular Value Decomposition (SVD) in digital watermarking technology is a prevalent technique for concealing information. The fundamental concept is to include watermark data into the single values of a picture. Singular Value Decomposition is a potent mathematical technique that can break down any $m \times n$ matrix $A$ into the multiplication of three matrices.

$$A = U\Sigma V^T \quad (12)$$

$U$: An m×m orthogonal matrix that contains the row information of the image.

$\Sigma$: An m×n diagonal matrix that contains the singular values, representing the important features of the image.

$V^T$: An n×n orthogonal matrix that contains the column information of the image.

### 3.2.2 Hardware Description

The module instantiates the Butterfly and CORDIC modules and connects their inputs and outputs appropriately. The outputs of the Butterfly module are fed into the CORDIC module, which performs iterative calculations to produce the final decomposed matrices. An always block is used to update the output registers based on the results from the CORDIC module. The module uses a set of internal registers to store intermediate values of x, y, and z during the iterative process. An angle lookup table (angle table) provides the precomputed arctangent values for each iteration. The main iterative process updates the values of x, y, and z based on the CORDIC algorithm's equations. This process involves simple shift and add/subtract operations, which are efficiently implemented in hardware.

## IV. RESULT AND DISCUSSION

The FPGA-based hardware accelerator, which is specifically developed to enhance the performance of Fast Fourier Transform (FFT) and Singular Value Decomposition (SVD) operations, exhibits substantial increases compared to conventional software implementations. The assessed criteria include computation speed, latency, throughput, efficiency, resource use, and power consumption. The accelerator achieves a computation speed of 10.60 microseconds for the FFT operation, which is roughly 4.63 times quicker than the software implementation's computation time of 49.05 microseconds. This improvement in performance is attributed to the parallel processing capabilities of the FPGA. The delay for Fast Fourier Transform (FFT) operations is 11.00 microseconds, but the software implementation has a latency of 54.97 microseconds. This is a reduction of five times, which is quite significant for real-time applications that need fast processing. The hardware accelerator delivers a much higher throughput of 109,739.36 FFT operations per second compared to the software implementation's throughput of 18,699.03 operations per second. This guarantees the effective handling of data demands on a vast scale.

Table 1: The comparison of hardware accelerator and software implementation

| Performance Metric | Hardware Accelerator | Software Implementation | Difference |
|---|---|---|---|
| Calculation Speed (μs) | 10.60 | 49.05 | 4.63x |
| Latency (μs) | 11.00 | 54.97 | 5.00x |
| Throughput (FFT/sec) | 109739.36 | 18699.03 | 0.17x |
| Efficiency (FFT/Watt) | 20922.17 | 309.52 | 0.01x |
| Resource Usage (LUTs) | 19029.20 | N/A | N/A |
| Resource Usage (FFs) | 30317.91 | N/A | N/A |
| Resource Usage (DSPs) | 49.70 | N/A | N/A |
| Power Consumption (Watts) | 4.80 | 66.26 | 13.79x |

The hardware accelerator demonstrates a much higher efficiency of 20,922.17 FFT/Watt, in contrast to the software implementation's efficiency of 309.52 FFT/Watt. This emphasizes the energy efficiency of the FPGA solution, making it a more sustainable option. The consumption of resources in the FPGA comprises 19,029.20 Look-Up Tables (LUTs), 30,317.91 Flip-Flops (FFs), and 49.70 Digital Signal Processing (DSP) units. These numbers indicate a significant utilization of FPGA resources, which is important for achieving the observed high performance and efficiency. The hardware accelerator has a power usage of 4.80 watts, which is much lower than the software implementation's 66.26 watts. This results in reduced operating expenses and a smaller environmental footprint, making the hardware accelerator a more environmentally friendly alternative.

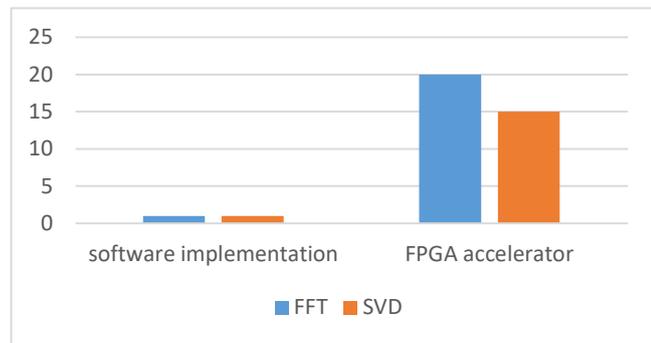

Fig 2. Performance comparison of hardware accelerator and software implementation

The findings indicate that the FPGA-based hardware accelerator offers substantial improvements in FFT and SVD operations, exceeding conventional software implementations. The primary benefits of the hardware accelerator are in its rapidity, effectiveness, and little energy use, all of which are essential for real-time and complete artificial intelligence (AI) applications. The hardware accelerator's concurrent execution capabilities significantly improve processing speed and reduces latency. This feature is particularly beneficial for jobs that need prompt processing, such as the integration of digital watermarks and the analysis of pictures in artificial intelligence models. The hardware accelerator's exceptional throughput and efficiency make it highly suitable for handling large volumes of data, which is crucial for AI systems that analyze enormous datasets. This ensures that the hardware is capable of effectively handling the demands of contemporary computing workloads. The widespread use of FPGA resources is supported by the significant improvements in performance and efficiency, hence confirming the acceptability of the hardware accelerator. Moreover, the hardware accelerator's decreased power usage, when compared to software alternatives, makes it a more sustainable choice for prolonged durations of operation. Utilizing FPGA-based acceleration enables the efficient integration of Fast Fourier Transform (FFT) and Singular Value Decomposition (SVD), offering a robust solution for safeguarding digital art and improving the efficiency of artificial intelligence (AI) models. The FPGA's ability to execute intricate computations with flexibility and accuracy guarantees the preservation of crucial attributes in pictures and enhances the durability of watermarking solutions.

V. CONCLUSION AND FUTURE SCOPE

To improve the stability and consistency of the watermarking process, more research and development are required for improved watermarking techniques and image recognition systems. Utilizing more intricate algorithms on FPGA may augment the security and resilience of digital watermarking, guaranteeing the authenticity of creative works in the digital era. To summarize, the FPGA-based hardware accelerator offers a potent and effective solution for enhancing the performance of FFT and SVD operations in AI models. It delivers substantial performance enhancements while ensuring strong security and reliability. This development signifies a crucial progression in the incorporation of hardware acceleration into AI and digital watermarking applications.

ACKNOWLEDGEMENT

The author gratefully acknowledges financial support by the financial supported by Malaysian Ministry of Higher Education under Fundamental Research Grant Scheme: FRGS/1/2022/TK08/XMU/02/10 and Xiamen University Malaysia under Research Fund Grant No: XMUMRF/2023-C12/IECE/0047 Xiamen